\documentclass[12pt]{article}
 \usepackage{epsfig}
 \def\be{\begin{equation}}
 \def\ee{\end{equation}}
 \def\bea{\begin{eqnarray}}
 \def\eea{\end{eqnarray}}
 \usepackage{graphicx}

 \catcode`\@=11
 \def\lsim{\mathrel{\mathpalette\@versim<}}
 \def\gsim{\mathrel{\mathpalette\@versim>}}
 \def\@versim#1#2{\vcenter{\offinterlineskip
 \ialign{$\m@th#1\hfil##\hfil$\crcr#2\crcr\sim\crcr } }}
 \catcode`\@=12

 \catcode`@=12
\begin{document}
 \thispagestyle{empty}
 \begin{flushright}
 UCRHEP-T605\\
 Dec 2020\
 \end{flushright}
 \vspace{0.6in}
 \begin{center}
 {\LARGE \bf Universal Scotogenic Fermion Masses\\ 
 in Left-Right Gauge Model\\}
 \vspace{1.2in}
 {\bf Ernest Ma\\}
 \vspace{0.2in}
{\sl Physics and Astronomy Department,\\ 
University of California, Riverside, California 92521, USA\\}
\end{center}
 \vspace{1.2in}

\begin{abstract}
In the conventional left-right gauge model, if the Higgs scalar sector 
consists only of an $SU(2)_L$ doublet and an $SU(2)_R$ doublet, fermion 
masses are zero at tree level.  There have been many studies on how 
they would become massive.  With the help of a dark sector with $U(1)_D$ 
gauge symmetry, it is shown how all standard-model fermions may acquire 
realistic masses radiatively, including that of the top quark.  In this 
context, the particle content of the model also implies the automatic 
conservation of baryon number $B$ and lepton number $L$ as in the 
standard model.  Observable anomalous Higgs couplings are predicted.
\end{abstract}

\newpage
\baselineskip 24pt
\noindent \underline{\it Introduction}~:~
In the standard $SU(3)_C \times SU(2)_L \times U(1)_Y$ gauge model (SM) of 
quarks and leptons, the one scalar Higgs doublet serves two purposes. 
It breaks $SU(2) \times U(1)_Y$ to the electromagnetic gauge symmetry 
$U(1)_Q$, and renders all fermions massive at tree level, except the 
left-handed doublet neutrino, unless it has a right-handed singlet 
counterpart.  In the canonical left-right extension to 
$SU(3)_C \times SU(2)_L \times SU(2)_R \times U(1)_{B-L}$, left-handed 
fermions are $SU(2)_L$ doublets, right-handed fermions are $SU(2)_R$ 
doublets.  The breaking of $SU(2)_L \times SU(2)_R \times U(1)_{B-L}$ to 
$U(1)_Q$ may be achieved by an $SU(2)_L$ Higgs doublet and an $SU(2)_R$ 
Higgs doublet, but they do not render the fermions massive at tree level. 
A scalar bidoublet under $SU(2)_L \times SU(2)_R$ would be needed, but 
is being withheld on purpose.   
There have been many studies on how quarks and leptons may acquire 
masses in this situation~\cite{dw87,m88,m89,bms03,gmr17,ms18,m20}.  They are 
usually not applicable to the top quark, because $m_t=173$ GeV is of order 
the electroweak breaking scale $v = \sqrt{2} \langle \phi^0 \rangle = 246$ 
GeV, and conventional wisdom would insist that it be accorded a tree-level 
mass~\cite{m90}.  In particular, if a radiative $m_t$ is desired, then it 
ought to be proportional to $v$, but suppressed by the typical loop factor of 
$16 \pi^2$.  Hence very large couplings to new particles are required 
and perturbative calculations become unreliable.  In this work, it will 
be shown how this objection may be overcome, and all SM fermion masses may  
be generated radiatively from a dark sector (scotogenic) with a $U(1)_D$ 
gauge symmetry in the left-right context.  Phenomenological consequences 
will be discussed.

\noindent \underline{\it Outline of Model}~:~
The particle content of the proposed model (similar to those of 
Refs.~\cite{gmr17,m20}) is listed in Table 1.
\begin{table}[tbh]
\centering
\begin{tabular}{|c|c|c|c|c|c|}
\hline
fermion/scalar & $SU(3)_C$ & $SU(2)_L$ & $SU(2)_R$ & $U(1)_{B-L}$ & $U(1)_D$ \\
\hline
$(u,d)_L$ & 3 & 2 & 1 & 1/6 & 0 \\ 
$(u,d)_R$ & 3 & 1 & 2 & 1/6 & 0 \\
$(\nu_e,e)_L$ & 1 & 2 & 1 & $-1/2$ & 0 \\ 
$(\nu_e,e)_R$ & 1 & 1 & 2 & $-1/2$ & 0 \\
\hline
$\Phi_L=(\phi_L^+,\phi_L^0)$ & 1 & 2 & 1 & 1/2 & 0 \\ 
$\Phi_R=(\phi_R^+,\phi_R^0)$ & 1 & 1 & 2 & 1/2 & 0 \\ 
\hline
$\zeta_L = (\zeta_L^{2/3},\zeta_L^{-1/3})$ & 3 & 2 & 1 & 1/6 & 1 \\ 
$\zeta_R = (\zeta_R^{2/3},\zeta_R^{-1/3})$ & 3 & 1 & 2 & 1/6 & 1 \\ 
$\eta_L=(\eta_L^0,\eta_L^-)$ & 1 & 2 & 1 & $-1/2$ & 1 \\ 
$\eta_R=(\eta_R^0,\eta_R^-)$ & 1 & 1 & 2 & $-1/2$ & 1 \\ 
\hline
$N_{L,R}$ & 1 & 1 & 1 & 0 & 1 \\
\hline
$\sigma$ & 1 & 1 & 1 & 0 & 3 \\
\hline
\end{tabular}
\caption{Fermion and scalar content of left-right model with $U(1)_D$.}
\end{table}
There are three families of quarks and leptons, as well as $N_{L,R}$. 
There is only one copy each of the scalars 
$\Phi_L,\Phi_R,\zeta_{L,R},\eta_{L,R},\sigma$.  The dark $U(1)_D$ gauge 
symmetry is broken by three units through the complex singlet scalar 
$\sigma$.  This allows a global $D$ symmetry to remain and prevents 
$N_{L,R}$ as well as $\nu_{L,R}$ to acquire Majorana masses, as pointed 
out first in Ref.~\cite{mpr13} and applied to Dirac neutrinos using $B-L$ 
in Ref.~\cite{ms15}.  From the allowed Yukawa couplings 
between the SM fermions and the dark particles, it is easily seen that 
the conventionally defined baryon number $B=1/3$ and $L=1$ for quarks 
and leptons are automatically transferred to $\zeta_{L,R}$ and $\eta_{L,R}$ 
with $N_{L,R}$ having $B=L=0$.

\noindent \underline{\it Origin of Large Radiative Top Mass}~:~
The one-loop diagram for a quark with charge $2/3$ is given in Fig.~1.
\begin{figure}[htb]
\vspace*{-5cm}
\hspace*{-3cm}
\includegraphics[scale=1.0]{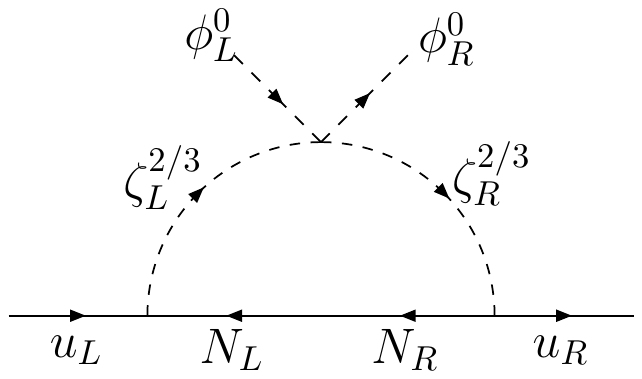}
\vspace*{-21.5cm}
\caption{Scotogenic $u$ quark mass.}
\end{figure}
The scalars $\zeta_{L,R}$ mix to form mass eigenstates
$\zeta_1 = \cos \theta \zeta_L - \sin \theta \zeta_R$, 
$\zeta_2 = \cos \theta \zeta_R + \sin \theta \zeta_L$, 
with masses $m_{1,2}$.  The diagram is then  easily calculated to be
\begin{equation}
m_u = {f_L f_R \sin \theta \cos \theta m_N \over 16 \pi^2} \left[ 
{m_2^2 \ln(m_2^2/m_N^2) \over 
m_2^2 - m_N^2} -  {m_1^2 \ln(m_1^2/m_N^2) \over m_1^2 - m_N^2} \right],
\end{equation}
which is of the same form as that of the original scotogenic model~\cite{m06} 
for Majorana neutrino mass.  The usual assumption is that $m_2^2-m_1^2$ is 
small compared to $m_0^2=(m_2^2+m_1^2)/2$ and $m_0 << m_N$, in which case 
a seesaw radiative mass is obtained, i.e.
\begin{equation}
m_u \simeq {f_L f_R \sin \theta \cos \theta (m_2^2-m_1^2) \over 16 \pi^2 m_N} 
\left[ \ln {m_N^2 \over m_0^2} - 1 \right].
\end{equation}
This clearly makes $m_u$ very small.  Another choice~\cite{gop10,m12} 
is $m_{1,2} >> m_N$, in which case
\begin{equation}
m_u \simeq {f_L f_R \sin \theta \cos \theta m_N \over 16 \pi^2} 
\ln {m_2^2 \over m_1^2}.
\end{equation}
This formula was applied~\cite{m12} to neutrinos where $m_N$ is of order 
keV to act as warm dark matter, but it is obvious that $m_N$ is actually 
arbitrary and may be chosen large enough to allow $m_t = 173$ GeV as a 
radiative effect.

Here the choice $m^2_2 >> m^2_{N_3} >> m^2_1$ is made, allowing different choices 
for $N_{1,2}$, to be discussed later.  Hence
\begin{equation}
m_t \simeq {f_L f_R \sin \theta \cos \theta m_{N_3} \over 16 \pi^2} 
\ln {m_2^2 \over m_{N_3}^2}.
\end{equation}
As an example, let $m_2 = 50$ TeV, $m_{N_3} = 15$ TeV, $m_1 = 1$ TeV, 
then $m_t = 173$ GeV is obtained for $f_L f_R \sin \theta \cos \theta = 0.682$.

The above may be achieved in the SM, but not in a very natural way.  First, 
the tree-level Higgs coupling to fermions must be forbidden by a new 
symmetry, say $Z_2$ under which all right-handed fermions are odd.  
This is often used for example in models 
where a small Dirac neutrino mass is desired~\cite{mp17}.  Another is to 
postulate a non-Abelian discrete family symmetry, such as $A_4$~\cite{mr01}, 
and assign left-handed and right-handed fermions differently so that they 
do not couple to the SM Higgs boson.  However, such symmetries must be 
softly broken appropriately to allow radiative fermion masses to 
appear~\cite{m14}.  To obtain $m_t=173$ GeV, this requires the 
corresponding soft breaking trilinear scalar term to have a coupling 
of order $10^7$ GeV.   In contrast, it is here 
naturally derived from the breaking of $SU(2)_R$ at a high scale.

\noindent \underline{\it Dark Scalar Sector}~:~
The $2 \times 2$ mass-squared matrix spaning $(\zeta_L,\zeta_R)$ is 
\begin{equation}
{\cal M}^2_\zeta = \pmatrix{m_L^2 & \lambda_{LR} v_L v_R \cr \lambda_{LR} 
v_L v_R & m_R^2},
\end{equation}
where the off-diagonal term comes from the quartic coupling 
$\lambda_{LR} (\zeta_L \phi^0_L)(\zeta_R \phi^0_R)^*$.  
Now $m_L^2 = m_1^2 \cos^2 \theta + m_2^2 \sin^2 \theta$, 
$m_R^2 = m_2^2 \cos^2 \theta + m_1^2 \sin^2 \theta$, and 
$\lambda_{LR} v_L v_R = (m_2^2 - m_1^2) \sin \theta \cos \theta$. 
This means that $v_R$ has to be very large for $m_2 = 50$ TeV, say of 
order $10^7$ GeV, so that the $SU(2)_R$ gauge bosons are out of reach 
at the Large Hadron Collider (LHC).

In the $d$ quark sector, the corresponding diagram is given in Fig.~2. 
\begin{figure}[htb]
\vspace*{-5cm}
\hspace*{-3cm}
\includegraphics[scale=1.0]{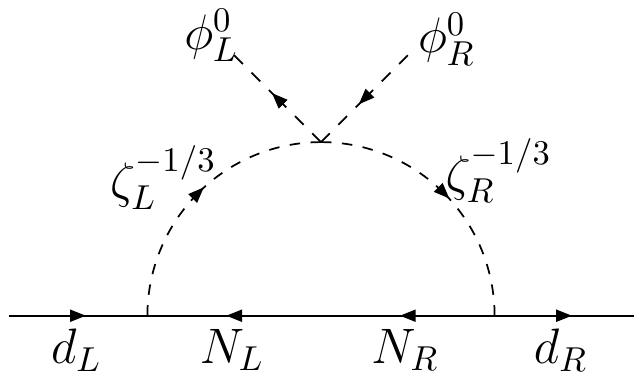}
\vspace*{-21.5cm}
\caption{Scotogenic $d$ quark mass.}
\end{figure}
Hence the off-diagonal entries of Eq.~(5) is replaced by $\lambda'_{LR}$ 
from the quartic coupling 
$\lambda'_{LR} (\zeta_L \bar{\phi^0}_L)(\zeta_R \bar{\phi}^0_R)^*$.  
This means that $m_{1,2}$ and $\theta$ in this sector may be  
different, allowing $m_b$ to be much smaller than $m_t$.

The lepton masses are obtained in exact analogy with the dark scalars 
$\eta_{L,R}$ instead of $\zeta_{L,R}$.  Whereas $N_{1,2,3}$ are common 
to quarks and leptons, the $\eta$ masses and their mixing angle in the 
charged-lepton and neutrino sectors are different.  The smallness of the 
Dirac neutrino masses are then related to the smallness of the quartic 
coupling $(\eta_L^0 \phi_L^0)(\eta_R^0 \phi_R^0)^*$.

\noindent \underline{\it Structure of Quark and Lepton Mass Matrices}~:~
Let the $N_{1,2,3}$ mass matrix be diagonalized, so that $m_{N_3} = 15$ 
TeV, $m_{N_2} = 800$ GeV, and $m_{N_1} = 10^8$ GeV.  Let the $3 \times 3$ 
quark mass matrix linking $(u,c,t)_L$ to $(u,c,t)_R$ be denoted by
\begin{equation}
{\cal M}_u = \pmatrix{m_{uu} & m_{uc} & m_{ut} \cr m_{cu} & m_{cc} & 
m_{ct} \cr m_{tu} & m_{tc} & m_{tt}}.
\end{equation} 
Whereas only $N_3$ contributes to $m_{tt}$ as in Eq.~(4), $N_2$ contributes 
to the $2 \times 2$ submatrix spanning $(c,t)$ of the form
\begin{equation}
m_{(c,t)} \simeq {f_L f_R \sin \theta \cos \theta m_{N_2} \over 16 \pi^2} 
\left[ \ln {m_2^2 \over m^2_{N_2}} - {m_1^2 \over m_1^2 - m^2_{N_2}} \ln 
{m_1^2 \over m^2_{N_2}} \right],
\end{equation}
and $N_1$ contributes to the entire $3 \times 3$ matrix of the form
\begin{equation}
m_{(u,c,t)} \simeq {f_L f_R \sin \theta \cos \theta m_2^2 \over 16 
\pi^2 m_{N_1}} \ln {m^2_{N_1} \over m_2^2}.
\end{equation}
It is clear that the $N_{3,2,1}$ contributions are of decreasing magnitude.
Similar structures appear in the $(d,s,b)$, $(e,\mu,\tau)$, and 
$(\nu_e,\nu_\mu,\nu_\tau)$ mass matrices.

\noindent \underline{\it Dark Matter}~:~
In this model, $N_2$ is the lightest particle of the dark sector.  It 
interacts mostly with the second family of quarks and leptons.  As such, 
its annihilation through $\zeta_1$ and $\eta_1$ to $c,s$ quarks and 
$\mu,\nu_\mu$ leptons may be large enough to have the correct relic 
abundance, and yet not affect the direct-search constraints which involve 
the $u,d$ quarks.  The  very heavy $N_1$ serves two purposes.  It allows 
very light masses for the first family of quarks and leptons; and it 
avoids any serious constraint from direct-search experiments.

The annihilation of $N_2 \bar{N}_2 \to c \bar{c}$ is dominated by $\zeta_1$, 
as shown in Fig.~3.
\begin{figure}[htb]
\vspace*{-5.5cm}
\hspace*{-3cm}
\includegraphics[scale=1.0]{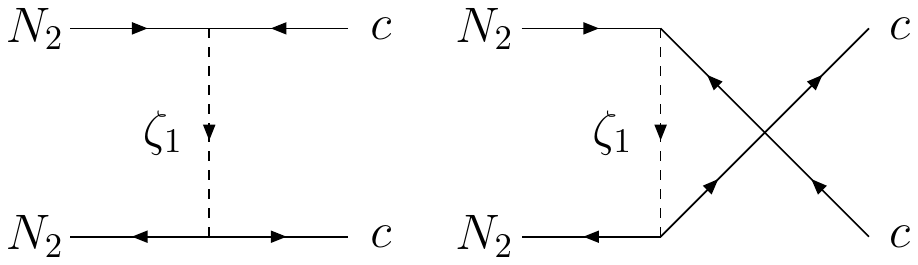}
\vspace*{-21.5cm}
\caption{Diagrams for $N_2 \bar{N}_2 \to c \bar{c}$.}
\end{figure}
The cross section $\times$ relative velocity is given by
\begin{equation}
\sigma_{ann} \times v_{rel} = {m_N^2 (f_L^4 \cos^4 \theta + 
f_R^4 \sin^4 \theta) \over 16 \pi (m_N^2 + m_1^2)^2}.
\end{equation}
This should be multiplied by a factor of 8 to account for the 3 colors 
of $c$ plus those of $s$, as well as $\mu$ and $\nu_\mu$, assuming these 
other contributions are the same in magnitude, and set equal to the 
canonical value of $3 \times 10^{-26}~{\rm cm}^3/{\rm s}$.  
For $m_{N_2} = 800$ GeV and $m_1 = 1$ TeV, 
\begin{equation}
f^4_L \cos^4 \theta + f^4_R \sin^4 \theta = 0.0678
\end{equation}
is obtained.  As an example, $\sin \theta = \cos \theta = 1/\sqrt{2}$ 
yields $(f_L^4+ f_R^4)^{1/4} = 0.72$.  Since $\zeta_1$ is analogous to 
a scalar quark in supersymmetry and $N_2$ analogous to a neutralino, 
they are subject to search limits at the LHC.  The updated ATLAS 
result~\cite{atlas20} shows that for $m_1 = 1$ TeV, $m_{N_2} = 800$ GeV 
is just at the edge of the allowed region.

\noindent \underline{\it Higgs and Gauge Sectors}~:~
The Higgs sector consists of the scalars $\Phi_{L,R}$ and $\sigma$.  Their 
potential is given by
\begin{eqnarray}
V &=& -\mu_L^2 \Phi_L^\dagger \Phi_L -\mu_R^2 \Phi_R^\dagger \Phi_R 
-\mu_\sigma^2 \sigma^* \sigma + {1 \over 2} \lambda_L (\Phi_L^\dagger 
\Phi_L)^2 + {1 \over 2} \lambda_R (\Phi_R^\dagger \Phi_R)^2 \nonumber \\ 
&+& {1 \over 2} \lambda_\sigma (\sigma^* \sigma)^2 + \lambda_{LR} 
(\Phi_L^\dagger \Phi_L)(\Phi_R^\dagger \Phi_R) + \lambda_{L\sigma}
(\Phi_L^\dagger \Phi_L)(\sigma^* \sigma) + \lambda_{R\sigma} 
(\Phi_R^\dagger \Phi_R)(\sigma^* \sigma).
\end{eqnarray}
After the spontaneous breaking of 
$SU(2)_L \times SU(2)_R \times U(1)_{B-L} \times U(1)_D$, the only physical 
scalars left are the real parts of $\phi^0_{L,R}$ and $\sigma$.  Let
\begin{equation}
\Phi_L = \pmatrix{0 \cr (v_L + h_L)/\sqrt{2}}, ~~~ \Phi_R = \pmatrix{0 \cr 
(v_R + h_R)/\sqrt{2}}, ~~~ \sigma = {1 \over \sqrt{2}} (v_D + h_D),
\end{equation}
then the $3 \times 3$ mass-squared matrix spanning $(h_L,h_R,h_D)$ is
\begin{equation}
{\cal M}^2_h = \pmatrix{\lambda_L v_L^2 & \lambda_{LR} v_L v_R & 
\lambda_{L\sigma} v_L v_D \cr \lambda_{LR} v_L v_R & \lambda_R v_R^2 & 
\lambda_{R\sigma} v_R v_D \cr \lambda_{L\sigma} v_L v_D & \lambda_{R\sigma} 
v_R v_D & \lambda_\sigma v_D^2}.
\end{equation}
Since $v_R \sim 10^{7}$ GeV, and $v_D$ should be at least a few TeV, $h_L$ 
acts to all intents and purposes as the SM Higgs boson in this model.
The heavier scalars $h_R$ and $h_D$ decay quickly to $h_L h_L$ through 
$\lambda_{LR}$ and $\lambda_{L\sigma}$ respectively.

In the gauge sector, the $Z_D$ boson gets a mass equal to $3 g_D v_D$. 
It should be heavier than $2m_{N_2}$, so it decays at least to 
$N_2 \bar{N}_2$.  The charged $W^\pm_{L,R}$ masses are $g_L v_L$ and 
$g_R v_R$.  The $Z,Z'$ mass-squared matrix is 
\begin{equation}
{\cal M}^2_{Z,Z'} = e^2 \pmatrix{v_L^2/x(1-x) & v_L^2/(1-x)\sqrt{1-2x} \cr 
v_L^2/(1-x)\sqrt{1-2x} & (1-x)v_R^2/x(1-2x) + xv_L^2/(1-x)(1-2x)},
\end{equation}
where $e^{-2} = g_L^{-2} + g_R^{-2} + g_B^{-2}$, and $g_L=g_R$ with 
$x=\sin^2 \theta_W$.  The $Z-Z'$ mixing is then about 
$x\sqrt{1-2x} v_L^2/(1-x)^2 v_R^2$, which is of order $10^{-10}$, much 
less than the experimental bound of about $10^{-4}$~\cite{pdg18}. 

\noindent \underline{\it Anomalous Higgs Couplings}~:~
In the SM, the Higgs boson $h$ has couplings to fermions fixed at $m_f/v$, 
where $v=246$ GeV.  Here $h_L$ has anomalous couplings, as first pointed 
out in Ref.~\cite{fm14}.  There are three contributions, the first being 
the $\lambda_{LR} (\zeta_L \phi_L^0)(\zeta_R \phi_R^0)^*$ coupling, the 
others from $|\phi_L^0|^2 |\zeta_L|^2$ and  
$|\phi_L^0|^2 |\zeta_R|^2$.  The latter are suppressed 
by the ratio $v_L/v_R$, and will be neglected.  The $h_L$ coupling to 
$\bar{t}t$ is then given by~\cite{fm14}
\begin{eqnarray}
&f_t& = {f_L f_R \sin \theta \cos \theta m_N \over 16 \pi^2 v_L} 
[(\cos^4 \theta + \sin^4 \theta)\left[ {m_2^2 \ln(m_2^2/m_N^2) \over 
m_2^2-m_N^2} - {m_1^2 \ln(m_1^2/m_N^2) \over m_1^2-m_N^2} \right] + \\ 
&& \sin^2 \theta \cos^2 \theta (m_2^2-m_1^2) \left[ 
{1 \over m_2^2 - m_N^2} - {m_N^2 \ln(m_2^2/m_N^2) \over (m_2^2-m_N^2)^2} + 
{1 \over m_1^2 - m_N^2} - {m_N^2 \ln(m_1^2/m_N^2) \over (m_1^2-m_N^2)^2} 
\right] ]. \nonumber 
\end{eqnarray}
Using the expression for $m_t$ from Eq.~(1), 
\begin{eqnarray}
f_t &=& {m_t \over v_L} [ \cos^4 \theta + \sin^4 \theta + \sin^2 \theta 
\cos^2 \theta (m_2^2-m_1^2) \left[ {m_2^2 \ln(m_2^2/m_N^2) \over 
m_2^2-m_N^2} - {m_1^2 \ln(m_1^2/m_N^2) \over m_1^2-m_N^2} \right]^{-1} 
\nonumber \\ ~~~ && ~~~~~~~ \left[ 
{1 \over m_2^2 - m_N^2} - {m_N^2 \ln(m_2^2/m_N^2) \over (m_2^2-m_N^2)^2} + 
{1 \over m_1^2 - m_N^2} - {m_N^2 \ln(m_1^2/m_N^2) \over (m_1^2-m_N^2)^2} 
\right]] \nonumber \\ 
&=& {m_t \over v_L} [ 1 + 17.213 \sin^2 \theta \cos^2 \theta ],
\end{eqnarray}
where the example of $m_N = 15$ TeV, $m_1 = 1$ TeV, and $m_2 = 50$ TeV has 
been used as previously.  Thus $f_t$ is predicted to be greater than that of 
the SM, which is a generic result~\cite{fm14}.  If $\sin \theta = 0.1$ is 
assumed, then the ratio of $f_t$ to that of the SM is 1.17.  The above may 
also be applied to $f_b$ with same $m_{L,R}$ but different $\theta$.  The 
present measurements of Higgs production and decay from collider 
data~\cite{pdg18} are consistent with the predictions of the SM, but there 
is much room for possible deviations.  A comprehensive analysis is left 
to future work.

\noindent \underline{\it Other Phenomenological Consequences}~:~
Each $3 \times 3$ radiative fermion mass matrix may be diagonalized by 
unitary matrices $U_L^\dagger$ on the left and $U_R$ on the right. 
However, they do not diagonalize the corresponding magnetic-moment 
matrix.  This means that on top of the contributions from the dark 
scalars to the anomalous magnetic moments of quarks and leptons, there 
are new sources of off-diagonal radiative transitions, such as 
$b \to s \gamma$, $\mu \to e \gamma$, etc.  Constraints from experimental 
data require these to be small.  A comprehensive analysis is left to 
future work.

Neutrinos are Dirac fermions in this model.  They have radiative masses, 
but are only suppressed relative to those of the charged leptons by 
making $\sin \theta$ in the neutrino sector very very small.  Thus the  
smallness of neutrino masses requires fine tuning, which is indeed a 
shortcoming of this model.  To remedy this situation, an $SU(2)_R$ Higgs 
triplet $(\Delta_R^{++},\Delta_R^+,\Delta_R^0)$ may be added to break 
$SU(2)_R$ at a high scale, so that $\nu_R$ gets a large Majorana mass. 
Lepton number $L$ now becomes lepton parity $(-1)^L$~\cite{m15}.  The usual 
seesaw mechanism applies and neutrinos obtain naturally small Majorana 
masses.  Note that $\Delta_R$ does not affect the generic radiative mechanism 
for the Dirac fermion masses. 

\noindent \underline{\it Conclusion}~:~
In the context of left-right gauge symmetry, a natural scenario exists 
where all fermions obtain radiative masses.  This is enforced by a very 
simple Higgs sector, consisting of one $SU(2)_L$ doublet and one $SU(2)_R$ 
doublet.  The absence of a scalar bidoublet means that all quarks and 
leptons are massless at tree level.  A dark sector is then assumed with 
a gauge $U(1)_D$ symmetry as shown in Table 1.  It is spontaneously 
broken by a complex singlet scalar with $D=3$.  The resulting theory 
conserves global $D$, as well as global baryon number $B$ and lepton 
number $L$.

The $SU(2)_R$ gauge symmetry is broken at a high scale, say of order 
$10^7$ GeV, allowing $m_t = 173$ GeV to be radiatively generated 
perturbatively in one loop.  This goes against the conventional wisdom 
that $m_t$ must be a tree-level mass, based on knowing that the electroweak 
breaking scale is $v_L = 246$ GeV.  Anomalous Higgs couplings 
to all fermions are predicted.  This will affect both the production 
and decay of the observed 125 GeV Higgs boson at the LHC.  It accentuates 
the importance of measuring all Higgs properties precisely in the future.

Dark matter is now intimately related to fermion mass generation.  It is 
a gauge singlet Dirac fermion $N_2$.  It couples mainly to the second 
family of quarks and leptons.  It annihilates through dark scalar exchange 
to $c,s,\mu,\nu_\mu$, but does not interact with nuclei significantly. 
Because of the dark color scalars which may be produced copiously in pairs 
at the LHC, $N_2$ may be observed as large missing momentum in their 
decays.

\noindent \underline{\it Acknowledgement}~:~
This work was supported in part by the U.~S.~Department of Energy Grant 
No. DE-SC0008541.

\bibliographystyle{unsrt}

\end{document}